\documentclass{ws-ijmpcs}

\usepackage{graphicx}
\usepackage{amssymb}
\usepackage{amsmath}
\input{epsf}
\usepackage{cite}

\def\@fmsl@sh#1#2#3{\m@th\ooalign{$\hfil#1\mkern#2/\hfil$\crcr$#1#3$}}
 \def\eq#1\en{\begin{equation}#1\end{equation}}
\def\s[#1,#2]{[#1\stackrel{\star}{,}#2]}
\def\sx[#1,#2]{[#1\stackrel{\star_{x}}{,}#2]}

\newcommand{\nc}{\newcommand}
\nc{\beq}{\begin{equation}}
\nc{\eeq}{\end{equation}}
\nc{\beqa}{\begin{eqnarray}}
\nc{\eeqa}{\end{eqnarray}}

\def\bc{\begin{center}}
\def\ec{\end{center}}

\def\to{\rightarrow}

\def\gsim{\mathrel{\mathpalette\atversim>}}

\def\bc{\begin{center}}
\def\ec{\end{center}}

\def\gsim{\mathrel{\rlap{\lower4pt\hbox{\hskip1pt$\sim$}}

    \raise1pt\hbox{$>$}}}       

\def\gsim{\mathrel{\rlap{\lower4pt\hbox{\hskip1pt$\sim$}}
    \raise1pt\hbox{$>$}}}       
\begin{document}

\makeatletter
\def\fmslash{\@ifnextchar[{\fmsl@sh}{\fmsl@sh[0mu]}}
\def\fmsl@sh[#1]#2{%
  \mathchoice
    {\@fmsl@sh\displaystyle{#1}{#2}}%
    {\@fmsl@sh\textstyle{#1}{#2}}%
    {\@fmsl@sh\scriptstyle{#1}{#2}}%
    {\@fmsl@sh\scriptscriptstyle{#1}{#2}}}
\def\@fmsl@sh#1#2#3{\m@th\ooalign{$\hfil#1\mkern#2/\hfil$\crcr$#1#3$}}
\makeatother

\markboth{S. M. Barr and X. Calmet}
{Grand Unification without Higgs Bosons}

%
\catchline{}{}{}{}{}
%

\title{Grand Unification without Higgs Bosons
}

\author{Stephen M. Barr}

\address{Department of Physics and Astronomy and the Bartol Research Institute \\
University of Delaware
\\ Newark,  Delaware 19716, USA.\\
smbarr@bartol.udel.edu}

\author{Xavier Calmet}

\address{Department of Physics $\&$ Astronomy \\
University of Sussex  \\ Falmer, Brighton, BN1 9QH, UK.\\
x.calmet@sussex.ac.uk}

\maketitle

\begin{history}
\received{Day Month Year}
\revised{Day Month Year}
\end{history}

\begin{abstract}
We discuss how a model for the electroweak interactions without a Higgs could be embedded into a grand unified theory. The requirement of a non-trivial fixed point in the $SU(2)$ sector of the weak interactions together with the requirement of the numerical unification of the gauge couplings leads to a prediction for the value of the $SU(2)$ gauge coupling in the fixed point regime. The fixed point regime must be in the TeV region to solve the unitarity problem in the elastic scattering of W bosons. We find that the unification scale is at about $10^{14}$ GeV. Viable grand unified theories must thus conserve baryon number. We discuss how to build such a model without using Higgs bosons.

\end{abstract}


This paper is dedicated to the memory of Julius Wess. While Julius
was a mathematical physicist, he truly had an amazing physical
intuition and understanding for experimental developments. He was
very much interested in seeing how his ideas could be applied to
particle physics. It turns out that several of his mathematical
constructions can be used to build elegant extensions of the
standard model of particle physics. The most well known example is
that of supersymmetry which is one of the most popular framework to
address the fine tuning problem of the Higgs boson's mass in the
standard model. This article will be dealing with another potential
extension of the standard model, without a Higgs boson this time,
using techniques developed in the 1960's by Callan, Coleman, Wess
and Zumino \cite{Callan:1969sn} who understood that gauge symmetries
could be implemented in a non-linear way. Nowadays, these models are
called non-linear sigma models and have found beautiful applications
in low energy QCD or in models of physics beyond the standard model
with a  composite Higgs boson.

Despite strong evidence for the existence of a standard model Higgs
coming from ATLAS and CMS at the Large Hadron Collider, it is
worthwhile to entertain the possibility that there is no fundamental
Higgs boson as experimental data is not conclusive yet and the about
3 $\sigma$ signal for a 125 GeV Higgs boson could turn out to be a
rather large background fluctuation.  Also from a purely academic
point of view, one can ask how one would build a Grand Unified
Theory without fundamental scalars. We shall first review briefly
how to build the standard model without a propagating Higgs boson
\cite{Calmet:2010ze,Calmet:2010cb} and then extend these ideas to
grand unified theories.

While the Higgs mechanism is an elegant tool to  generate masses for
the electroweak gauge bosons while preserving the perturbative
unitarity of the S-matrix and the renormalizability of the theory,
it is well understood that gauge symmetries can be implemented in a
non-linear way  \cite{Callan:1969sn}. We shall emphasize that mass terms for the electroweak
bosons can be written in a gauge invariant way using a non-linear
sigma model representation. We start with the Lagrangian
\begin{eqnarray}
\label{eq:sofar}
 L_{SM}=  L_0 + L_{mass}
\end{eqnarray}
where
\begin{eqnarray}
L_0 =  -\frac{1}{4} W^a_{\mu\nu} W^{a\, \mu\nu} -\frac{1}{4} B_{\mu\nu} B^{\mu\nu} - \frac{1}{4} G_{\mu\nu} G^{\mu\nu}
 + \sum_{j=1}^3 \,  \bar\Psi^{(j)} i\fmslash{D} \Psi^{(j)}
 \end{eqnarray}
 and
 \begin{eqnarray}
 L_{mass} &=& M_W^2\, W^+_{\mu} W^{-\, \mu} + \frac{1}{2} M_Z^2\, Z^\mu Z_\mu \\  \nonumber  &&
  \qquad\qquad - \sum_{i,j} \Big( \bar u^{(i)}_L M^u_{ij} u_R^{(j)} + \bar d^{(i)}_L M^d_{ij} d_R^{(j)}
     + \bar e^{(i)}_L M^e_{ij} e_R^{(j)} + \bar \nu^{(i)}_L M^\nu_{ij} \nu_R^{(j)} \Big)
  + h.c. \, ,
\end{eqnarray}
where $\Psi =\{ q^i_L , u^i_R , d^i_R , l^i_L , e^i_R , \nu^i_R \}$
stands for the standard model fermions and $i,j$ are generation
indices. The mass terms for the gauge bosons and fermions above arise from the manifestly gauge invariant Lagrangian 
\begin{equation}
\label{eq:chiralLmass}
 L_{mass} = \frac{v^2}{4} \text{Tr}\left[ \left( D_\mu \Sigma \right)^\dagger
 \left( D^\mu \Sigma \right) \right] - \frac{v}{\sqrt{2}} \sum_{i,j} \left( \bar u_L^{(i)} d_L^{(i)} \right)
 \Sigma \begin{pmatrix} \lambda_{ij}^u\,  u_R^{(j)} \\[0.1cm] \lambda_{ij}^d\,  d_R^{(j)} \end{pmatrix} + h.c.
\end{equation}
The field $\Sigma(x)$ is given by
 \begin{eqnarray}
\Sigma(x) = \exp(i\sigma^a \chi^a(x)/v)
\end{eqnarray}
where $\chi_a$ are the Goldstone bosons of $SU(2)_L$ and the covariant derivative by
 \begin{eqnarray}
D_\mu \Sigma = \partial_\mu \Sigma 
 -i g \frac{\sigma^a}{2} W^a_\mu \Sigma + i g^\prime \Sigma \frac{\sigma^3}{2} {\cal A}_\mu \, .
\end{eqnarray}
The local $SU(2)_L\times U(1)_Y$ invariance is evident, since
$\Sigma$ transforms as
\begin{eqnarray}
\Sigma \to U_L(x) \, \Sigma\, U_Y^\dagger(x)
\end{eqnarray}
with
\begin{eqnarray}
  U_L(x) = \exp \big(i\alpha_L^a(x) \sigma^a/2 \big)  \ \mbox{and} \
  U_Y(x) = \exp \big(i\alpha_Y(x) \sigma^3/2 \big).
\end{eqnarray}
The fields $\chi^a$ are transforming in a non-linear way under a
local $SU(2)_L\times U(1)_Y$  gauge transformations
\begin{eqnarray}
\chi^a(x) \to \chi^a(x) + \frac{v}{2}\, \alpha_L^a(x) - \frac{v}{2}\,\delta^{a3}\, \alpha_Y(x) \, .
\end{eqnarray}

In the unitary gauge, $\langle \Sigma \rangle = 1$, the chiral
Lagrangian (\ref{eq:chiralLmass}) reproduces the mass term of
Eq.(\ref{eq:sofar}) with
\begin{equation}
\label{eq:rho}
\rho \equiv \frac{M_W^2}{M_Z^2 \cos^2\theta_W} = 1 \, .
\end{equation}
This relation is consistent with the experimentally measured value to quite good accuracy.
It follows as the consequence of  the approximate invariance of (\ref{eq:chiralLmass})
under  $SU(2)_L\times SU(2)_R$ global transformations,
\begin{equation}
\Sigma \to U_L \, \Sigma\, U_R^\dagger \, ,
\end{equation}
which is spontaneously broken to the diagonal subgroup  $SU(2)_c$ by $\langle \Sigma \rangle = 1$, and
explicitly broken by $g^\prime\not =0$ and $\lambda^u_{ij} \not = \lambda^d_{ij}$.
In the limit of vanishing $g^\prime$ the fields $\chi^a$ transform as a triplet under the
``custodial'' $SU(2)_c$, so that $M_W = M_Z$.

This formulation is mathematically identical to the one presented in
terms of gauge invariant fields in \cite{Calmet:2010cb}. The
electroweak bosons can be defined in terms of  gauge invariant
fields given by
 \begin{eqnarray}
\underline W^i_\mu&=& \frac{i}{2g} \mbox{Tr} \ \Omega^\dagger \stackrel{\leftrightarrow}
{D_\mu} \Omega \tau^i
\end{eqnarray}
with $D_\mu=\partial_\mu - i  g  \frac{\sigma_a}{2} B^a_\mu(x)$ and
  \begin{eqnarray}
\Omega=\frac{1}{\sqrt{\phi^\dagger
    \phi}}\left(\begin{array}{cc}  \phi_2^* & \phi_1
    \\ -\phi_1^* & \phi_2
  \end{array}
\right )
\end{eqnarray}
 where
 \begin{eqnarray}
\phi=\left(\begin{array}{c}  \phi_1 \\
   \phi_2
  \end{array}
\right ).
\end{eqnarray}
is a $SU(2)_L$ doublet scalar field which is considered to be a
dressing field and does not need to propagate.  The very same
construction can be applied to fermions
\cite{Calmet:2010cb,tHooft:1998pk,'tHooft:1980xb,Mack:1977xu,Visnjic:1987pj}:
\begin{eqnarray}
\underline \psi^a_L&=&\Omega^\dagger \psi^a_L
\end{eqnarray}
where $ \psi^a_L$ are the usual $SU(2)_L$ doublets.

Now clearly the standard model without a propagating Higgs boson is
not renormalizable at the perturbative level. In fact, an infinite
number of higher dimensional operators will be generated by
radiation corrections:
  \begin{eqnarray} \label{effaction}
S=S_{SM w/o Higgs}+\int d^4 x  \sum_i \frac{C_i}{v^n} O^{4+n}_i
\end{eqnarray}
 where $O^{4+n}_i$ are operators compatible with the symmetries of the model.
 There is a beautiful analogy between this theory and quantum general relativity
 \begin{eqnarray} \label{effaction2}
S[g]= -\int d^4x \sqrt{-\det(g)}\! \! \! \! \! && \left   (-
\Lambda(\mu) +\frac{\bar M_P^2(\mu)}{32 \pi} R+  a(\mu) R_{\mu\nu}
R^{\mu\nu} +b(\mu) R^2 \right . \\ \nonumber && \left .
+\frac{c(\mu)}{\bar M_P^2} R^3+\frac{d(\mu)}{\bar M_P^2} R
R_{\mu\nu} R^{\mu\nu}+ .... \right ).
\end{eqnarray}
While quantum general relativity is not renormalizable at the
perturbative level as well, Weinberg \cite{fixedpoint} argued some
40 years ago that it might be renormalizable at the non perturbative
level if there is a non-trivial fixed point. Evidence for the
existence of such a non -trivial fixed point has accumulated over the
last 40 years. This approach is known as asymptotic safety. While
calculations to identify such fixed points are difficult and often
involve uncontrolled approximation, this approach to quantum gravity
is clearly very exciting as it is very conservative and does not
require new speculative theories to formulate quantum gravity.

If there is no propagating Higgs boson, it is most natural to posit
that the weak interactions are asymptotically safe. This implies the
existence of non-trivial fixed points in the $SU(2)_L$ gauge
interactions and Yukawa sectors. Amazingly, there are indeed
indications of a non-trivial fixed point in non linear sigma models
\cite{Fabbrichesi:2010xy,Percacci:2009dt,Percacci:2009fh}. It should
be stressed nevertheless that these indications rely on a truncation
of the effective action which can be considered as an uncontrolled
approximation. However, while in the case of gravity it might never
possible to observe the fixed point regime, we expect that this will
be the case at the LHC for the weak interactions, if this mechanism
has been chosen by nature.

Besides a non-trivial fixed point, the weak scale must have a
non-trivial running to suppress the growth of the W-W scattering
amplitude with energy.  Formally, one can introduce  a dimensionless
coupling constant:
\begin{eqnarray}
g_{v} = k^2/v(k)^2
\end{eqnarray}
where $k$ is the renormalization group momentum scale. A
renormalization group equation for $g_v$ can be written as
\begin{eqnarray}
\frac{\partial g_v}{\partial \ln k}= \beta_k = (d-2+ \eta) g_v
\end{eqnarray}
where $d$ is the number of dimensions and with the anomalous dimension of the weak bosons given by
\begin{eqnarray}
\eta = - \frac{\partial \ln Z(k) }{\partial \ln k}
\end{eqnarray}
which in general is a function of all the couplings of the
Lagrangian (\ref{effaction}). $Z(k)$ denotes the wave-function
renormalisation factor of the electroweak bosons. In four
dimensions, there is a non-trivial fixed point if $\eta=-2$. In
analogy to the non-perturbative running of the non-perturbative
Planck mass,  we introduce an effective weak scale
 \begin{eqnarray}
v_{eff}^2=v^2\left(1+\frac{\omega}{8\pi} \frac{\mu^2}{v^2} \right)
\end{eqnarray}
where $\mu$ is some arbitrary mass scale, $\omega$ a
non-perturbative parameter which determines the running of the
effective weak scale and $v$ is the weak scale measured at low
energies. If $\omega$ is positive, the electroweak interactions
would become weaker with increasing center of mass energy.
Suppressing the growth of the partial wave amplitude enough to
maintain the unitarity bound at all scales requires $\omega\ge 1/3$.

While the model formulated above represents a perfectly natural and
consistent ultraviolet completion of the standard model, it is
natural to wonder whether grand unification is compatible with that
framework. Grand unified theories are  fascinating models and the
realization that the fermions of the standard model fit naturally
into representations of simple groups such as $SU(5)$ or $SO(10)$ is
a strong motivation to consider these ideas. As always, the messy
part comes with the spontaneous symmetry breaking of these large
group to that of the standard model. Realistic models require the
inclusion of  many new Higgs bosons which has always led to
theoretical problems such as the hierarchy problem, the
doublet-triplet mass-splitting problem, and higher dimensional
operators involving the Higgs bosons which can lead to sizable
threshold effects which are very difficult to evaluate.  It is thus
tempting to build grand unified theories without Higgs bosons while
preserving the beautiful aspects of grand unification.

\begin{figure}
\centering
\includegraphics[width=3in]{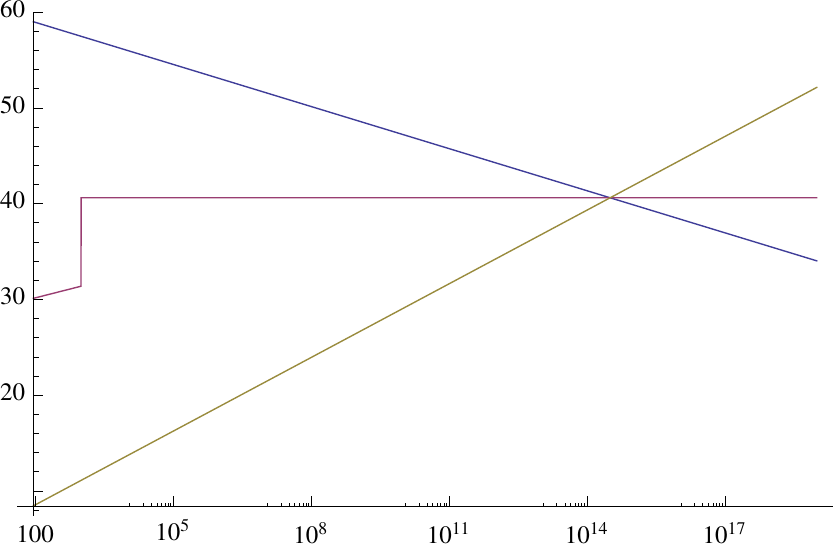}
\caption{This is a logarithmic plot of $\alpha_i^{-1}$ as a function of the running scale $\mu$.}
\label{Figure1}
\end{figure}

The requirement of the numerical unification of the coupling of the
standard model without a Higgs boson leads to an interesting
prediction for the value of the $SU(2)$ gauge coupling in the fixed
point regime. If there is a non-trivial fixed point, the beta
function of the $SU(2)_L$ is zero and the coupling constant is thus
fixed. The beta functions of the $U(1)$ and $SU(3)$ are given by
their usual Standard Model values with the caveat that there are no
scalars and the value of the $U(1)$ beta function in our model will
thus differ from that of the standard model. We have
 \begin{eqnarray}
 \mu \frac{\partial}{\partial \mu} \alpha_i(\mu) = \frac{1}{2 \pi} b_i \alpha_i^2(\mu)
\end{eqnarray}
with
 \begin{eqnarray}
 b^{UV}_i=\left (\begin{array}{c} b_1\\b_2\\b_3\end{array} \right)=
 \left (\begin{array}{c} 0\\ 0\\-11\end{array} \right)+ N_{gen} \left (\begin{array}{c} 4/3\\0\\4/3\end{array} \right)
 \end{eqnarray}

 we assume that the number of generations $N_{gen}=3$. Requiring
 that the $U(1)$ and $SU(3)$ gauge coupling meet in one point allow
 us to determine the unified coupling constant and the unification
 scale. We find that the unified coupling constant is $\alpha_\star=1/40.6$ and
 the unification scale $\mu_\star=3.2 \times 10^{14}$ GeV. This allows us to
 determine the value of the coupling constant in the fixed point domain
 which is identical to the unified coupling constant. This is a
 prediction of this unification mechanism. Our result is illustrated
 in Fig. (\ref{Figure1}). We assumed that the fixed point is in
 the TeV region. This is necessary to suppress the growth of
 the W-W scattering amplitude. Below that scale we used the
 usual standard model value for the beta function of the $SU(2)$
 however without a Higgs doublet, i.e. $b_2=-22/3+N_{gen} 4/3$.
 
Note that in Fig. (\ref{Figure1}), we assume that the running of the $SU(2)_L$ coupling constant is given by the standard model equation without a Higgs boson and replaced by the fixed-point value above a TeV. This approximation is similar to what is done in supersymmetric models where it is assumed that the superpartners do not contribute below one TeV, while they do above the energy scale above which supersymmetry is restore. In our case, we expect the running to be more complicated than this close to the fixed-point. We expect that the $SU(2)_L$ coupling constant will increase, hence making the higher dimensional operators discussed above more important before reaching its constant value in the fixed-point region. We stress that our approximation is not important for our numerical evaluation.

An obvious issue is that if we use conventional grand unified
theories such as $SU(5)$\cite{Georgi:1974sy} or
$SO(10)$\cite{Fritzsch:1974nn}, there are dimension six operators
which lead to a proton decay rate which is excluded by current
observations. We are thus forced to consider grand unified theories
which do not have proton decay. Fortunately, such models do exist
\cite{Berezhiani:2001ub,Kobakhidze:2001yk}. The reason for proton
decay in most grand unified models is that leptons and quarks are
unified in the same multiplets of the grand unified theory. For
example, in $SU(5)$ they are unified in the $\bar 5$ and $10$, the
heavy gauge bosons $X$ and $Y$ of $SU(5)$ with masses of the order
of the unification scale lead to a fast decay of the proton. There
are several possibilities to suppress proton decay if the model is
formulated in higher dimensions
\cite{Berezhiani:2001ub,Kobakhidze:2001yk} (see also \cite{Langacker:1977ai,Langacker:1978fn} for older GUT models with a stable proton). In four dimensions one
can rely on more complicated unified structures such as $SU(3)^3$
trinification or $SU(N) \times SU(N)$ which typically do not have
$X$ and $Y$ gauge bosons and thus no gauge-mediated proton decay if
the matter representations are chosen appropriately. While in
conventional models with a Higgs mechanism there can still be proton
decay mediated by colored Higgs bosons, we would not have such a
problem with our approach where the Higgs bosons are replaced by non
linear sigma models.

It is also possible to have models with simple groups
\cite{Berezhiani:2001ub}, for example a $SU(5)$ grand unified theory
with an unusual fermion assignment. The authors of
\cite{Berezhiani:2001ub}, for example, suggest taking states $\bar
F_{1,2}\sim \bar 5$ and $T_{1,2,3}\sim 10$ and anti-states $F$ and
$\bar T_{1,2}$ per generation.  The $SU(5)$ symmetry is broken by
the fields $\Phi$ and $\Omega$, both in the reducible 24+1
representations, having the following vacuum expectations: $\langle
\Phi \rangle= v_1  \mbox{diag}(1, 1, 1, 0, 0)$ and $\langle \Omega
\rangle= v_2  \mbox{diag}(0, 0, 0, 1, 1)$ for which in our approach
we represent by non linear sigma models. They  also introduce one
extra 24 + 1 field $\Sigma$ which develops a vacuum expectation
value of the form $\langle  \Sigma \rangle= \sigma \mbox{diag}(1, 1,
1, -1, -1)$.  A Yukawa interaction
 \begin{eqnarray}
(\bar F_1 F + T_1 \bar T_1) \Phi +(\bar F_2 F +T_2 \bar T_1) \Omega + T_3 \bar T_2 \Sigma
 \end{eqnarray}
where $\Phi$, $\Omega$ and $\Sigma$, in our approach, are non-linear
sigma models will give masses to the heavy fermions, the light
fermions of the standard model can be identified as in
\cite{Berezhiani:2001ub}: $L \subset \bar F_1$, $d^c \subset \bar
F_2$, $e^c \subset T_1$, $u^c \subset T_2$ and $Q \subset T_3$. This
assignment insures that lepton and baryon numbers are preserved to
all orders in perturbation theory.

In conclusion, we have discussed how a model for the electroweak
interactions without a Higgs could be embedded into a grand unified
theory. The requirement of a non-trivial fixed point in the $SU(2)$
sector of the weak interactions together with the requirement of the
numerical unification of the gauge couplings leads to a prediction
for the value of the $SU(2)$ gauge coupling in the fixed point
regime. The fixed point regime must be in the TeV region to solve
the unitarity problem in the elastic scattering of W bosons. We find
that the unification scale is at about $10^{14}$ GeV. Viable grand
unified theories must thus conserve baryon number. We have discussed
how to build such a model without using Higgs bosons.


\end{document}